\documentstyle[eqsecnum,aps]{revtex}
\newcommand{\gex}{gedankenexperiment}

\newcommand{\cA}{{\cal A}}

\newcommand{\deri}[1]{ {d \over d #1} }

\newcommand{\Einf}{{E_\infty}}

\newcommand{\Finf}{{F_\infty}}

\newcommand{\Winf}{{W_\infty}}
\newcommand{\DSG}{{\Delta S_G}}
\newcommand{\DSBH}{{\Delta S_{BH}}}

\newcommand{\TBH}{{T_{BH}}}

\newcommand{\rrad}{{\rho_{rad}}}
\newcommand{\Prad}{{P_{rad}}}
\newcommand{\srad}{{s_{rad}}}
\newcommand{\vel}{{\vec {e_l}}}
\newcommand{\vk}{{\vec {k}}}
\newcommand{\vn}{{\vec {n}}}
\newcommand{\vx}{{\vec {x}}}
\newcommand{\vP}{{\vec {P}}}
\def\rslash{\partial\kern-0.026em\raise0.17ex\llap{/}%
\kern0.026em\relax}
\def\Dslash{D\kern-0.15em\raise0.17ex\llap{/}\kern0.15em\relax}
\def\sqr#1#2{{\vcenter{\hrule height.#2pt
      \hbox{\vrule width.#2pt height#1pt \kern#1pt
          \vrule width.#2pt}
      \hrule height.#2pt}}}
\def\square{{\mathchoice{\sqr63}{\sqr63}{\sqr63}{\sqr63}}}
\begin{document}
\draft
\title{%
\hspace*{14.5cm} {\rm OCHA-PP-158}\\
\vspace{0.5cm}
	A Comment on Non-Archimedean Character of Quantum Buoyancy
}
\author{%
	Takashi Okamura%
	\footnote{E-mail: okamura@skymint.phys.ocha.ac.jp}
}
\address{%
	Department of Physics, Ochanomizu University, \\%
	1-1 Otsuka, 2 Bunkyo, Tokyo, 112-8610, Japan
}
\date{\today}
\maketitle
\begin{abstract}
In a {\it \gex } about the generalized second law (GSL)
of black hole thermodynamics,
the buoyant force by black hole atmosphere (the acceleration radiation)
plays an important role, and then it is significant to understand
the nature of the buoyant force.
Recently, Bekenstein criticizes that the fluid approximation of
the acceleration radiation which is often used in the estimation
of the buoyant force is invalid for the case that
the size of the target is much less than a typical wavelength
of the acceleration radiation,
due to the diffractive effect of wave scattering.
He calculated the buoyant force as a wave scattering process and
found that the buoyant force as a wave scattering process is weaker
than in the fluid approximation.
And he asserts that while the buoyant force by black hole atmosphere
is insufficient for the GSL to hold, the Bekenstein's entropy bound
is enough.
In this letter, we argue that even if it is correct that 
we should calculate the buoyant force as a wave scattering process,
its implication in the GSL strongly depends on whether there exists
any massless scalar field, that is, S-wave scattering.
By reconsidering the diffractive effect by S-wave scattering,
we show that if some massless scalar field exists,
then the GSL can hold without invoking a new physics,
such as an entropy bound for matter.
\end{abstract}
\pacs{PACS number(s): 04.70.Dy, 04.70.Bw, 05.30.-d, 97.60.Lf}
\section{Introduction}
We believe that black hole is a thermodynamical object
and has entropy in a sense.
This belief is based on analogy with ordinary thermodynamics;
mathematical relationship between black hole mechanics
\cite{BardeenETAL73}
and ordinary law of thermodynamics, and
the existence of thermal radiation from a black hole.

In order to transmute the mathematical relationship into physical one,
that is, black hole mechanics into black hole {\it thermodynamics},
understanding of statistical origin of black hole entropy
has progressed from the various points of view
\cite{stat}.
It is any theory of quantum gravity which controls
a measure of the density of the states that we need to understand
genuinely statistical black hole entropy without divergent quantities.
On the other hand, black hole entropy can be derived
in metrical theories of gravity by a classical method
such as Noether charge method
\cite{Wald93}
or by semiclassical methods such as Euclidean path integral method
\cite{GH77},
though these methods by no means count quantum degrees of freedom
that are responsible for black hole entropy.
Since any successful quantum gravity theory comes to a corresponding
metrical theory of gravity in suitable low energy limit,
we expect that the value of black hole entropy obtained by classical or
semiclassical methods should be also derived by
counting quantum degrees of freedom.
Thus, we may regard success of statistical derivation
of black hole entropy as a benchmark test of
a proposed quantum gravity theory.

On the other hand, it is also necessary to understand
black hole entropy better, even at the level of thermodynamics.
For instance, it is the second law that characterizes
the peculiar property of entropy in ordinary thermodynamics
because of its referring to a direction in time.
Therefore it is very important for understanding of black hole entropy
to establish the second law of black hole thermodynamics.

Since we cannot regard a black hole as an isolated system owing to
the universal interaction with ordinary matter outside the black hole
by gravity, any second law of black hole thermodynamics should refer to
total entropy of self-gravitating system including black holes.
Therefore we are led to the generalized second law (GSL)
of black hole thermodynamics, which asserts that in any process,
the generalized entropy
\begin{equation}
	S_G:= S_{BH}+S_M = {1 \over \TBH} \left( 
	{\kappa A_{BH} \over 8 \pi} \right) + S_M
\label{eq:GSL}
\end{equation}
never decreases, where $S_{BH}$ and $S_M$ denote the entropy
of the black holes and that of ordinary matter outside the black hole,
and then $\kappa$, $A_{BH}$ and $\TBH$ are the surface gravity, area of
the event horizon and temperature of the black hole, respectively%
\footnote{%
Someone may doubt additivity of entropy of self-gravitating system,
due to long-range nature of gravity.
Instead of no conclusive argument about the additivity,
we assume the validity of the additivity.
See \cite{MY89} for argument validating the additivity.
}.
The validity of the GSL is essential for the consistency of black hole
thermodynamics and for the interpretation of the horizon area
as representing the physical entropy of a black hole,
because it is nothing but the ordinary second law for self-gravitating
systems containing black holes.
Thus, the GSL is a cornerstone of black hole thermodynamics.

Although an explicit general proof of the GSL has not been given
until now, the various attempts for special cases have been performed
\cite{Bek81,UW82,UW83,Zurek82,Sorkin86,FP93,Myers94,FiolaETAL}.
Considering a process which transfers an infinitesimal energy $\delta E$ 
and entropy $\delta S$ in the external region into the black hole
adiabatically, we obtain the change in the total entropy
$\delta S_G=\delta E/\TBH-\delta S$.

In classical theory, we may argue as follows.
Since a black hole can classically export nothing outside the horizon,
it is natural to give zero temperature $\TBH=0$ to the black hole.
Therefore, if it were so, by dominance of the first term
in Eq.(\ref{eq:GSL}), the GSL in classical theory should be
no more than the second law for the black hole entropy alone
and then it would amount to the area increasing law
in black hole mechanics $\delta A_{BH}>0$ which holds
by energy condition $\delta E>0$
\cite{Haw71}.

However, it is awkward to assign $\TBH=0$ to the black hole,
since the black hole entropy or the change in it becomes divergent
and ill-defined.
Thus, the physical analogy appears end in classical theory.
In order to have non-zero black hole temperature and
well-defined black hole entropy, 
it is indeed essential to incorporate quantum effects
even semiclassically.
Due to the breakdown of the energy condition of quantum fields,
black holes can radiate and acquire non-zero temperature,
and then the thermodynamic quantities of a black hole can be fixed as
$\TBH=\kappa/2\pi$ and $S_{BH}=A_{BH}/4$
\cite{Haw7475}.
Therefore, it is important to investigate the validity of the GSL
by consistent arguments with taking account of quantum effects.

An observer accelerating with acceleration $a$ detects
isotropic thermal radiation with temperature $T_U=\hbar a/2 \pi$
by the Unruh radiation (acceleration radiation)
\cite{Unruh76}.
An object suspended near a black hole is accelerated
by virtue of its being prevented from following a geodesic.
Unruh and Wald
\cite{UW82,UW83}
suggested that this object will likewise see Unruh radiance.
Since its acceleration ({\it i.e.} temperature) varies
with distance from the horizon, they surmised that the object
will be subject to a buoyant force by the acceleration radiation
{\it fluid} and the buoyancy affects the energetics of a process
which exchange entropy and energy between the black hole and
outer matter.
They concluded that quantum buoyancy is sufficient by itself
to protect the GSL.

Recently, Bekenstein reconsidered the nature of acceleration radiation
and its implication on the GSL
\cite{Bek99}.
He pointed out that the wave nature (diffractive effect) of
the acceleration radiation cannot be neglected in the case that
the size of the object lowered toward the black hole is smaller than
a typical wavelength of the acceleration radiation and that
the fluid approximation of the acceleration radiation is invalid.
For such a case, he estimated the buoyant force
as a wave scattering process and found that the buoyant force
as a wave scattering process is weaker than in the fluid approximation.
Therefore, the diffractive effect alters energetics of exchange process
of the entropy and energy compared with that in fluid picture,
and then the quantum buoyancy is insufficient by itself
to protect the GSL.
A breakdown of the GSL in the existing physics leads us
to a new physics, such as an entropy bound for matter,
if we take granted that the GSL holds.
Thus, the question of the validity of the GSL is still be opened
even in a simple \gex.

In this letter, we observe that if a massless scalar field exists,
the quantum buoyancy is sufficient to protect the GSL,
rather strengthen the validity of the GSL, even though
we take account of the wave nature of the acceleration radiation.
\section{a gedankenexperiment}\label{Sec:review}
In this section, we specify a \gex~investigated in this letter
and review two independent reasonings for the GSL to hold.

We consider a static black hole which area of the event horizon
is $\cA$ and a box of proper height $b$ and
geometrical cross-sectional area $A$.
Far from the black hole, the box is filled with matter,
so that the total energy of the box and contents
is $E_0$ and its total entropy $S_0$.
Subsequently, the box is lowered adiabatically toward the black hole
by a weightless rope to some height $l$ that is the proper distance
between the horizon and the center of mass of the box.
And then, the box and contents are released and dropped into
the black hole.

Because of the process to be adiabatic,
the total entropy of the box and contents remains constant.
Therefore, the change in the total entropy becomes%
\footnote{%
Here we implicitly assume that processes after the box released
preserves, the total energy and entropy of the box and contents.
}
\begin{equation}
 \DSG = \DSBH - S_0 = {\Delta M(l) \over \TBH} - S_0~,
\label{eq:DSGI}
\end{equation}
where $\TBH$ is the (non-zero) black hole temperature.

On the other hand, the energy $\Delta M(l)$
delivered to the black hole decreases during the lowering process,
because the gravitational energy of the box and contents
is lost by the work against the tension of the rope.
As denoting the redshift factor $\xi(l)$, we obtain the equation,
\begin{eqnarray}
 \Delta M(l) &=& E_0 + (\mbox{work done by the rope}) = E_0 + \Winf(l)
  = E_0 + \int^l_\infty \left( - \Finf^G \right)~ dl
\label{eq:DMI} \\
 &=& E_0 + E_0 \left[~ \xi(l) - 1~ \right] = E_0~ \xi(l)~,
\label{eq:DMII}
\end{eqnarray}
where we use the relation
$-\Finf^G~dl = d \Einf = d \left( E_0 \xi \right)$,
that is derived from $\Einf = E_0~ \xi$.
Thus, the energy delivered to the black hole is
\lq\lq redshifted away", due to the negative gravitational potential.

Therefore, the change in the total entropy is
\begin{equation}
 \DSG={E_0 \over \TBH}~ \xi(l) - S_0 = {E_0 \over T(l)} - S_0~,
\label{eq:DSGII}
\end{equation}
where $T(l):=\TBH/\xi(l)$ means the locally measured temperature
of the black hole atmosphere.
Because, if the box can be close to the horizon without limit,
$\xi$ can be arbitrary small near the horizon,
we can make the value of $\DSG$ negative at will.

If we take granted that the GSL holds, then
we need any mechanism which prevents the box from the horizon.
At present, there exist two reasonings: one is invoking to
an entropy bound for matter and the other makes use of
the buoyant force by the black hole atmosphere.
It is essential to recognize that the box must have a finite size
which is greater than its Compton wavelength.

The argument of the first reasoning invoking an entropy bound
is as follows:
The finiteness of the box size imposes a constraint,
$l \ge b/2$, that is,
\begin{equation}
	\xi(l) \sim \kappa~l = 2 \pi \TBH l \ge \pi \TBH~ b~,
\label{eq:preBekbound}
\end{equation}
because the bottom of the box cannot touch the horizon.
If we premise the validity of the GSL and
the energetics Eq.(\ref{eq:DMII}),
we need an entropy bound for matter
\cite{Bek81}
\begin{equation}
 S \le \pi E~ b~.
\label{eq:Bekbound}
\end{equation}
Thus, the entropy of any matter in this case
is bounded above by its energy and size.
Since, obviously, the size $b/2$ is greater than
its gravitational radius $r_g=2E$, we obtain
\begin{equation}
 S \le 2 \pi E~ r_g \le {4 \pi r_g^2 \over 4}~.
\label{eq:BekboundII}
\end{equation}
Thus the maximum entropy of any matter is bounded above by
its gravitational radius and the saturated state
is attained by the black hole state.
This relation is called holographic bound,
which the validity of Eq.(\ref{eq:BekboundII}) is open problem and
has actively been discussed in the different viewpoint,
holographic principle
\cite{holography}.
Even though it is finally true that there exists the entropy bound
for matter or the holographic bound, it is important to investigate
to what degree the GSL is protected by the known physics and
whether the validity of the GSL implies the entropy bound or
the holographic bound.

Another reasoning invoking the known physics makes use of
quantum effect of matter field outside black holes, that is,
the buoyant force by the black hole atmosphere, which has been neglected
in the argument of the first reasoning.
We may start with two main working hypothesis
\cite{UW82,UW83,PW99};
\begin{description}
 \item[A1.] The black hole atmosphere is describable
	by radiation {\it fluid} of a unconstrained thermal matter
	which is defined to be the state of matter that maximizes
	entropy density at a fixed energy density and
	the radiation {\it fluid} has the locally measured temperature $T(l)$.
\item[A2.] The buoyant force on the box exerted
	by the black hole atmosphere is equal to the pressure gradient of
	the radiation {\it fluid} of unconstrained thermal matter.
\end{description}
The assumption {\bf A1} means that the Gibbs-Duhem relation holds,
\begin{equation}
 \cases{ \rrad + \Prad - T(l)~ \srad = 0 \cr
          d \rrad = T(l)~d\srad \cr}~,
\label{eq:assumptionI}
\end{equation}
and by Eqs.(\ref{eq:assumptionI}) and $T(l)=\TBH/\xi(l)$,
we obtain balance equation between
gravitational force and pressure gradient force of the radiation fluid
\begin{equation}
	\deri{l}(\xi \Prad) = - \rrad(l)~ {d \xi \over dl}~.
\label{eq:pressuregrad}
\end{equation}
Using Eq.(\ref{eq:pressuregrad}) and the assumption {\bf A2}, we obtain
\lq\lq Archimedean principle",
\begin{equation}
 \Finf^B= A \left[~ (\xi \Prad)_{bottom}-(\xi \Prad)_{top}~ \right]
        =-V \deri{l}(\xi \Prad) =  V \rrad(l)~{d \xi \over dl}~.
\label{eq:Archimedean}
\end{equation}
Therefore, the work done by the total force $\Finf=\Finf^G + \Finf^B$
becomes
\begin{equation}
 \Winf=\int^l_\infty \left( - \Finf \right)~ dl
  = E_0 \left[~ \xi(l) - 1~ \right] + V~ \xi(l)~ \Prad(l)~.
\label{eq:totalwork}
\end{equation}
And then, the energy delivered into the black hole is
\begin{eqnarray}
 \Delta M &=& E_0+\Winf = \left[~ E_0+ V \Prad(l)~ \right]~ \xi(l)
\\
 &=& V \left[~ \rho_0 - \rrad + T(l)~\srad~ \right]~ \xi(l)~,
\label{eq:DMbuoyant}
\end{eqnarray}
where $\rho_0 := E_0/V$ is the average energy density
of the box and the contents.
The change in the total entropy becomes
\begin{equation}
	\DSG= V \left( {\rho_0 -\rrad \over T(l)} + \srad - s_0 \right)~,
\label{eq:DSGbuoyant}
\end{equation}
where $s_0 := S_0/V$ is the average entropy density
of the box and the contents.

The critical situation for the positivity of $\DSG$ is the case of
minimizing $\Delta M$,
\begin{equation}
  0 = \deri{l} \Winf = \Finf^G + \Finf^B
 = V \left(~ \rho_0 - \rrad~ \right)~ {d \xi \over dl}~,
\label{eq:minwork}
\end{equation}
so that, it is the most dangerous for the validity of the GSL
when the box is dropped into the black hole at the floating point
$\rho_0 = \rrad(l)$.

Nevertheless the positivity of $\DSG$ holds by the definition
of the radiation fluid, that is,
we can show the validity of the GSL
\cite{UW82,UW83,PW99}
without invoking a new physics,
\begin{equation}
	\DSG \ge V \left( \srad - s_0 \right) \ge 0~,
\label{eq:minDSG}
\end{equation}
where the last inequality follows the definition {\bf A1}
of the radiation fluid, because of $\rho_0=\rrad(l)$
at the floating point.

Now we should check the validity of our assumptions,
especially, the validity of the assumption {\bf A2}.
It is natural to think that if a typical wavelength
of the acceleration radiation $\lambda$ is much bigger than
the box size $b$, the assumption {\bf A2} is invalid
due to the breakdown of the fluid picture, such as diffractive effect.
Therefore, it is doubtful to consider that the fluid picture is still
valid far from the black hole,
such as $b < \lambda \sim T^{-1}(l)$.
\section{The buoyant force by long wavelength scattering}
Recently, Bekenstein pointed out
the breakdown of the fluid picture far from the horizon
\cite{Bek99}.

Strictly speaking, the true pressure exerted on the surface of the box
is given by integrating true stress tensor over the surface
and the true stress tensor must be obtained by inclusion of
the boundary condition of the surface.
However, in the previous section, we estimated the pressure
by the fictitious stress tensor, which means that the stress tensor is
estimated by neglecting the surface,
exclusive of the boundary condition.
In order to estimate the true pressure, it is often useful
to calculate the {\it change} in momentum flux on the surface
and it is essential for calculating the change in the momentum flux
to estimate the reflection coefficient, that is,
to include the boundary condition on the surface.
For example, a perfectly transparent glass is not exerted by photons,
even though the momentum flux across the glass does not vanish.

Thus, we need to estimate the scattering cross section of the box
for the acceleration radiation.
If $b \ll R_H$, where $R_H$ is the curvature radius at the horizon,
then we can acquire a large local (Lorentz) frame including
the target (the box) in which the target is at rest.
Therefore, we can approximate the scattering process in the black hole
spacetime by the scattering process in flat spacetime and
at first estimate quantities in interest, such as the momentum transfer,
in the local frame.
A remained task is to transform quantities obtained in the local frame
into ones in the global frame, that is, quantities as measured
at infinity
\cite{Bek99}.

We calculate physical quantities in the long wavelength limit
$b \ll \lambda=:2 \pi/k$, because we are especially interested in
scattering phenomena in the situation that the fluid picture
of the acceleration radiation is suspicious.
In this limit, the differential cross section is indifferent
to details such as the shape of the target.
Hereafter we assume that the the shape of the target is
spherically symmetric.
\subsection{the buoyant force by dipole scattering}
According to the above procedure, Bekenstein estimated the buoyant
force exerted by dipole scattering.
In order to estimate the buoyant force, we calculate
the differential cross-section of the target object with the size $b$
by the dipole scattering.
For the dipole scattering which transfers the incident wave
with the wave vector $\vk$ into the scattered wave with $\vk'$
and preserves the magnitude of the momentum, $k:=|\vk|=|\vk'|$,
we have
\begin{equation}
 {d \sigma \over d \Omega'} =
  b^2~\left(~ kb~ \right)^4 F(\vn,~\vn')~,
\label{eq:dipolecross}
\end{equation}
where $\vn$ and $\vn'$ are a pair of the unit vectors
denoting the incident and scattering directions, $\vn:=\vk/k$ and
$\vn':=\vk'/k$, respectively.
And $F$ is some dimensionless function, which, for example,
is given by $F(\vn, \vn')= \pi^{-2}~\left\{ {5 \over 8}
\left[ 1+ \cos^2 \left( \vn \cdot \vn' \right) \right]
- \cos \left( \vn \cdot \vn' \right) \right\}$ for electromagnetic
scattering from a conducting sphere.
The fourth order dependence of the cross-section on the wave vector
is attributed to the fact that the dipole part is dominant
in the scattering of the electromagnetic wave.

Given a distribution function of the incident wave as
$f(k)=1/\left[~\exp\left( \hbar k / T \right)-1~\right]$,
the incident momentum flux carried in the acceleration radiation
in the vicinity of the wave vector $\vk=k \vn$ becomes
\begin{eqnarray}
 \vn~ I(k,\vn)~ dk d\vn
  &:=& (\hbar \vk)~ f(k)~ {d^3k \over (2\pi)^3}
  = \vn~ \hbar k^3~ f(k)~ {dk d\vn \over (2\pi)^3}~,
\label{eq:incmomflux}
\end{eqnarray}
where $d^3k = k^2 dk~d\vn$.

Because the fraction $d\sigma/d\Omega'$ among the incident flux
$I~ dk d\vn$ is scattered into the direction $\vn'$,
we obtain the momentum transfer of the box in the local frame,
\begin{eqnarray}
 {d\vP \over d\tau}=\int dk d\vn \int d\vn'~
  I(k,\vn)~ {d\sigma \over d\Omega'}(\vn - \vn')~,
\label{eq:momtranI}
\end{eqnarray}
where $\tau$ is time measured in the local frame, that is,
the proper time of the target.

Since the acceleration radiation has the temperature gradient,
the radiation going to the direction $\vn$ hits on
the box with the local temperature
$T(l,\vn)=\TBH / \xi(l,\vn) = \hbar/2 \pi [l+(\vel \cdot \vn) b]
=T(l)/[1+(\vel \cdot \vn) b/l]$,
where $\vel$ is the unit vector that is directed to the center of
mass of the box from the black hole.
Therefore, the buoyant force $\Finf^{Scatt}$ by the dipole scattering
in the global frame becomes
\begin{equation}
 \Finf^{Scatt} = \xi(l) \left\vert {d\vP \over d\tau} \right\vert \sim
	(2 \pi l~ \TBH) \left[{T(l) \over \hbar} \right]^8 b^6
	\int d\vn \left[ 1+{b \over l} (\vel \cdot \vn) \right]^{-8}
	\sim {\TBH \over b} \left(~ {b \over l}~ \right)^8~,
\label{eq:BFfordipole}
\end{equation}
where we neglect numerical factor.
Thus, the buoyant force by the dipole scattering is proportional to
the seventh power of the size $b$, not to the volume (non-Archimedean)
and proportional to the eighth inverse power of the proper distance $l$
from the event horizon.

On the other hand, the buoyant force in fluid picture is estimated by,
\begin{equation}
 \Finf^B =  V \rrad(l)~ {d \xi \over dl} 
 \sim b^3~ [T(l)]^4~\TBH
 \sim {\TBH \over b}~\left(~ {b \over l}~ \right)^4~.
\label{eq BFinfluid}
\end{equation}
Thus, in the case of the dipole scattering, Archimedean character
of buoyant force that the force is proportional to
the volume of the box does not work for $b \ll \lambda$ and
the force rapidly decreases with the distance from the horizon
than in the fluid picture.

Since it is possible to saturate the inequality Eq.(\ref{eq:minDSG})
by lowering the radiation matter, even in the case
for the fluid picture to be valid,
the fact that the buoyant force by dipole scattering is weaker than
in the fluid picture suggests that buoyant force alone is not enough
for the GSL to be valid.
Indeed, we can show that there exists cases satisfying both of
the breakdown of the GSL and the validity of the approximation used
\cite{Bek99}.
\subsection{the buoyant force by S-wave scattering}
The non-Archimedean character of buoyant force shown
in the previous subsection is attributed to
the dipole dominant scattering.
If we assume that a massless%
\footnote{%
Since we concentrate on the case that the lowering process goes on
far from the black hole in order to make the used approximations valid,
massive fields less contribute to buoyant force.
The reason is that massive quanta far from the black hole are
much less \lq\lq excited" for a stationary observer than massless ones.
}
scalar field exists in nature,
the argument based on dipole scattering does not work and
implication on the GSL by wave nature of the acceleration radiation
is drastically changed, due to S-wave scattering.

By mode decomposition of the equation of motion of the scalar field
$\square~ \phi=0$ with respect to the plane wave in the local frame,
we obtain
\begin{eqnarray}
	& & \left[~k^2 + \Delta~ \right]~ \Psi_\vk=0~, \hspace{1cm}
	\phi={\exp(-i k \tau) \over \sqrt{2 k}}
	{\Psi_\vk(\vx) \over (2\pi)^{3/2}}~.
\label{eq:modedecom}
\end{eqnarray}

Since we would like to consider the case that
the total entropy of the box and contents remains constant,
we regard the surface of the box as infinite potential barrier.
Therefore, we solve a scattering problem by
the infinite potential barrier at the radius $b$ in quantum mechanics.
We easily obtain the result
\cite{Messiah},
\begin{eqnarray}
 & & \Psi_\vk(\vx) \sim \exp(i \vk \cdot \vx) + g(\Omega){\exp(ikr) \over r}~;~
 \hspace{1cm} {d\sigma \over d\Omega}=|g(\Omega)|^2~,
\label{eq:bcatinfty} \\
 & &g(\Omega) = \sum^\infty_{l=0}{2l+1 \over 2ik}
  \left( 1+{h^{(2)}_l(kb) \over h^{(1)}_l(kb)} \right) P_l(\cos\theta)~,
\label{eq:ampofscattwave} 
\end{eqnarray}
where $h^{(n)}_l$ and $P_l$ are the spherical Hankel function
of the $n$-th kind and the Legendre one of the first kind, respectively.
In the long wavelength limit, the reflection coefficient $g(\Omega)$
is approximated by
\begin{eqnarray}
	& &g(\Omega) \sim -b \sum^\infty_{l=0}{(kb)^{2l} \over [(2l-1)!!]^2}
	P_l(\cos\theta)~.
\label{eq:ampofscattwaveL} 
\end{eqnarray}

If S-wave scattering occurs,
we have the differential cross section independent of wavelength,
\begin{equation}
	\left(~{d\sigma \over d\Omega}~\right)_{l=0}
	\sim b^2 \sim {\sigma_T \over 4\pi}~, \hspace{0.5cm}
	\sigma_T \sim 4\pi~ b^2  = 4 A~,
\label{eq:diffcrosssecS}
\end{equation}
where $\sigma_T$ is the total scattering cross section of the target.

In this connection, if the dipole dominated scattering occurs,
such as the electromagnetic field,
we have the differential cross section
\begin{equation}
	\left(~{d\sigma \over d\Omega}~\right)_{l=1}
	\sim b^2 (kb)^4 \left[~ P_1(\cos\theta)~ \right]^2~,
\label{eq:diffcrosssecP}
\end{equation}
which depends on the fourth power of $k$ as Eq.(\ref{eq:dipolecross}).

The contribution of the S-wave scattering to
the momentum transfer of the target in the local frame is
\begin{eqnarray}
 & &{d\vP \over d\tau}=\int dk d\vn \int d\vn'~
  I(k,\vn)~ \left( {d\sigma \over d\Omega'} \right)_{l=0}(\vn - \vn')
  = \int d\vn~ \sigma_T~ \int dk \left[~\vn~I(k,\vn)~ \right]~,
\label{eq:momtranS}
\end{eqnarray}
where the last equality is due to the spherical symmetric scattering
of S-wave.
Since the quantity $\sigma_T~ \int dk \left[~\vn~I(k,\vn)~ \right]$
is nothing but the momentum flux across the surface with the area
$\sigma_T$ into the direction $\vn$,
Eq.(\ref{eq:momtranS}) gives momentum transfer four times
larger than the total momentum flux across the surface of the box.

Therefore, the buoyant force in the global frame exerted on the box
by S-wave scattering is four times larger than that in the fluid picture,
because of the diffractive effect,
\begin{eqnarray}
 & &\Finf^{Scatt} = 4 \Finf^B(\mbox{fluid})
  = 4 V \rrad(l)~ {d \xi \over dl}~.
\label{eq:BFofSwave}
\end{eqnarray}
Since the diffractive effect of S-wave scattering strengthens
the buoyant force than in the fluid picture,
the above fact suggests that the GSL is protected
by the buoyant force alone.

Indeed, following the argument in Sec.\ref{Sec:review} for this case,
we obtain the inequality from Eq.(\ref{eq:DSGbuoyant})
\begin{eqnarray}
	{\DSG \over V} & \ge & 3~{\rrad \over T} + \srad(\rrad) -s_0
\label{eq:DSGSI} \\
	 & \ge & 3~{\rrad \over T} + \srad(\rrad) - \srad( 4 \rrad )~,
\label{eq:DSGSII}
\end{eqnarray}
where we explicitly denote the dependency of $\srad$ on $\rrad$.
The first line is given by $\rho_0=4 \rrad$ at the floating point
and the second comes from the assumption {\bf A1},
$s_0 \le \srad(\rho_0)=\srad( 4 \rrad )$.
Using the equations,
\begin{eqnarray}
	& &\srad(\rrad) = {4 \over 3} {\rrad \over T}~,
\\
	& &{\srad( \rrad ) \over \srad( \rho'_{rad} ) } =
	\left( { \rrad \over \rho'_{rad}  }\right)^{3/4}~,
\end{eqnarray}
we can show the validity of the GSL
\begin{equation}
	{\DSG \over V} \ge \srad(\rrad) \left( {13 \over 4}
	- 2^{{3 \over 2}} \right) > 0~.
\label{eq:DSGSlast}
\end{equation}
Thus, if some massless scalar field exists, then
without invoking a new physics such as an entropy bound for matter,
the GSL holds thanks to the buoyant force strengthened by
the diffractive effect of S-wave scattering of black hole atmosphere.

For the completeness, we should check the validity of
the fluid picture in short wavelength limit.
In this limit, we obtain the differential cross section
\begin{eqnarray}
	& &{d\sigma \over d\Omega}
	 \sim {A \over 4 \pi}~,
\label{eq:crossforS}
\end{eqnarray}
and finally obtain the momentum transfer in the local frame as
\begin{eqnarray}
	{d\vP \over d\tau} &=& \int dk d\vn \int d\vn'~
	I(k,\vn)~ \left( {d\sigma \over d\Omega'} \right)(\vn - \vn')
	= \int dk d\vn~\left[ \vn~I(k,\vn) \right] \int d\vn'~{A \over 4 \pi}
\nonumber \\
	&=& \int d\vn~ A~ \int dk \left[~ \vn~I(k,\vn)~ \right]~.
\label{eq:momtranSII}
\end{eqnarray}
As expected, the buoyant force in the geometrical optics approximation
is equal to that in the fluid picture.
\section{Summary}
In this letter, we briefly reviewed a \gex~to test the validity of
the GSL, which is any process composed of adiabatically lowering
the object toward the black hole and dropping into.
In the analysis of this \gex, the buoyant force by the black hole atmosphere
plays an important role and the buoyant force is usually estimated
by the pressure gradient of the radiation {\it fluid}.
However, since the pressure exerted on the target is given by
the change in the momentum flux,
it is necessary to estimate the reflection coefficient on the surface
of the target, in order to get the correct buoyant force.
In the case that the size of the target $b$ is larger than a typical
wavelength of the black hole atmosphere $\lambda$,
the pressure exerted on the surface of the box is well estimated by
the fluid picture for the black hole atmosphere.
On the other hand, in the case that the lowering process goes on
with satisfying $b < \lambda$, we cannot complete the reasoning
which makes the GSL to hold by the buoyant force estimated
in based on the fluid picture, because the fluid picture breaks down
by diffractive effect of wave scattering.

For buoyant force far from the black hole, massless fields dominate
over massive ones, due to less acceleration of
the quasi-stationary target compared with their masses.
Furthermore, in the long wavelength limit, the dependence of
the scattering cross section on wavelength much varies
according to the spin of the scattered wave.
Therefore, it much depends on the spin of massless fields in nature
whether we need to invoke a new physics
such as an entropy bound for matter, in order to hold the GSL, or not.
If some massless scalar field exists in nature, then the GSL can hold,
due to the buoyant force alone by black hole atmosphere.
If not so, the validity of the GSL might suggest
the existence of some new physics such as an entropy bound.

The above conclusion is based on the viewpoint of
an accelerated observer who rest on the box lowering adiabatically.
Although the energy-momentum tensor normalized by
the accelerated observer is different from the true one,
we can expect that the calculation of buoyant force
by the viewpoint of the accelerated observer gives correct estimation.
It is because the essential quantity in the calculation is gradient of
the energy-momentum tensor, not value itself, and
the difference between the energy-momentum tensor normalized by
the accelerated observer and the true one is divergence free.

In the Ref.
\cite{UW82},
it was shown that in two dimensional spacetime,
the estimation of $\Delta M(l)$ delivered to the black hole
in an accelerating viewpoint with the fluid approximation
is equivalent to that in an inertial point of view.
Does this equivalence suggest that the estimation of buoyant force
by wave scattering is different from that in an inertial point of view,
that is, not physical?
Since, in two dimensional spacetime, the reflection coefficient of
wave scattering by infinite potential is unity,
two estimations of buoyant force in an accelerating viewpoint with
and without the fluid approximation are equal to one another.
Therefore, three estimations, including in an inertial point of view,
are consistent and this result is due to the peculiarity of
two dimensional spacetime.
For completeness, it is worthwhile to estimate energetics $\Delta M(l)$
from an inertial point of view for higher dimensional spacetimes.

Furthermore, although we regard the mere sum of black hole entropy
and matter one as the total entropy, we have not yet obtained
the foundation.
Since gravity is long range force, it may be doubtful to assume
the additivity of entropies of individual systems
in self-gravitating system.
It is future work to reconsider the GSL without the assumption
of the additivity of entropies
\cite{Sorkin86}.
\acknowledgments
The author would like to thank Professor T. Mishima
and Dr. T. Shimomura for a careful reading of the manuscript.
He also appreciate Professor M. Morikawa for kind hospitality
in Ochanomizu University.


\begin{thebibliography}{99}
\bibitem{BardeenETAL73} J.M.~Bardeen, B.~Carter and S.W.~Hawking, %
Commun. Math. Phys. {\bf 31}, 161 (1973).
\bibitem{stat}%
W.H.~Zurek and K.S.~Thorne, Phys. Rev. Lett. {\bf 54}, 2171 (1986)~;\\
L.~Bombelli, R.~Koul, J.~Lee, and R.~Sorkin, Phys. Rev. D {\bf 34} 373 (1986)~;\\
V.~Frolov and I.~Novikov, {\it ibid}. {\bf 48} (1993) 4545~;\\
M.~Srednicki, Phys. Rev. Lett. {\bf 71}, 666 (1993)~;\\
S.~Carlip, Phys. Rev. D {\bf 51}, 632 (1995)~;\\
A.~Strominger and C.~Vafa, Phys. Lett. B {\bf 379}, 99 (1996)~;\\
A.~Strominger, JHEP {\bf 02}, 009 (1998)~;\\
S.~Carlip, Phys. Rev. Lett. {\bf 82}, 2828 (1999)~;\\
S.~Carlip, Class. Quant. Grav. {\bf 16}, 3327 (1999).
\bibitem{Wald93} R.M.~Wald, Phys. Rev. D {\bf 48}, R3427 (1993).
\bibitem{GH77} G.~Gibbons and S.W.~Hawking, Phys. Rev. D {\bf 15}, 2752 (1977).
\bibitem{MY89} E.A.~Martinez and J.W.~York, Jr., Phys. Rev. D {\bf 40}, 2124 (1989).
\bibitem{Bek81} J.D.~Bekenstein, Phys. Rev. D {\bf 23}, 287 (1981).
\bibitem{UW82} W.G.~Unruh and R.M.~Wald, Phys. Rev. D {\bf 25}, 942 (1982).
\bibitem{UW83} W.G.~Unruh and R.M.~Wald, Phys. Rev. D {\bf 27}, 2271 (1983).
\bibitem{Zurek82}
	W.H.~Zurek, Phys. Rev. Lett. {\bf 49}, 1683 (1982).
\bibitem{Sorkin86}
	R.D.~Sorkin, Phys. Rev. Lett. {\bf 56}, 1885 (1986).
\bibitem{FP93}
	V.P.~Frolov and D.N.~Page, Phys. Rev. Lett. {\bf 71}, 3902 (1993).
\bibitem{Myers94}
	R.C.~Myers, Phys. Rev. D {\bf 50}, 6412 (1994).
\bibitem{FiolaETAL}
	T.M.~Fiola, J.~Preskill, A.~Strominger, and S.P.~Trivedi, Phys. Rev. D {\bf 50}, 3987 (1994)~;\\
	T.~Shimomura, T.~Okamura, T.~Mishima and H.~Ishihara, gr-qc/9902029 (1999),
	to appear in Phys. Rev. D.
\bibitem{Haw71} S.W.~Hawking, Phys. Rev. Lett. {\bf 26}, 1344 (1971).
\bibitem{Haw7475} S.W.~Hawking, Nature(London) {\bf 248}, 30 (1974)~;~
	Commun. Math. Phys. {\bf 43}, 199 (1975).
\bibitem{Unruh76} W.G.~Unruh, Phys. Rev. D {\bf 14}, 870 (1976).
\bibitem{Bek99} J.D.~Bekenstein, Phys. Rev. D {\bf 60}, 124010 (1999).
\bibitem{holography} G.~'tHooft, gr-qc/9310026 (1993)~;\\
L.~Susskind, J. Math. Phys. {\bf 36}, 6377~ (1995)~;\\
R.~Bousso, JHEP {\bf 06}, 078 (1999); {\bf 07}, 004 (1999).
\bibitem{Messiah} A.~Messiah, {\it Quantum Mechanics}
(Elsevier-North-Holland, Amsterdam, 1961).
\bibitem{PW99} M.A.~Pelath and R.M.~Wald, Phys. Rev. D {\bf 60}, 104009-1 (1999).
\end{thebibliography}
\end{document}